\documentclass[preprint,tightenlines,showpacs,amsmath,amssymb]{revtex4-1}
\usepackage{epsfig}
\usepackage{dcolumn}
\usepackage{bm}
\usepackage{color}
\usepackage{graphicx}
\usepackage{subfigure}
\usepackage{pstricks}
\usepackage{pst-node}
\usepackage{rotating}
\usepackage{times}
\usepackage{overpic}
\usepackage{lineno}

\newcommand{\chicJ}{\chi_{cJ}}

\newcommand{\ssb}{\Sigma^0\bar{\Sigma}^0}
\newcommand{\ccb}{c\bar{c}}
\newcommand{\SSB}{\Sigma^+\bar{\Sigma}^-}
\newcommand{\xxb}{\Xi^0\bar{\Xi^0}}

\newcommand{\pbar}{\bar{p}}

\newcommand{\EE}{e^+e^-}

\newcommand{\GG}{\gamma\gamma}

\newcommand{\BB}{B\bar{B}}
\newcommand{\ppb}{p\bar{p}}

\newcommand{\pppr}{\pi^+\pi^-p\bar{p}}

\newcommand{\psp}{\psi^{\prime}}
\newcommand{\jpsi}{J/\psi}
\newcommand{\ar}{\rightarrow}

\newcommand{\ppjpsi}{\pi^0\pi^0 J/\psi}

\newcommand{\llb}{\Lambda\bar{\Lambda}}

\newcommand{\lamb}{\bar{\Lambda}}

\newcommand{\bfg}{\begin{figure}}
\newcommand{\efg}{\end{figure}}
\newcommand{\bitm}{\begin{itemize}}
\newcommand{\eitm}{\end{itemize}}
\newcommand{\bnum}{\begin{enumerate}}
\newcommand{\enum}{\end{enumerate}}
\newcommand{\btbl}{\begin{table}}
\newcommand{\etbl}{\end{table}}
\newcommand{\btbu}{\begin{tabular}}
\newcommand{\etbu}{\end{tabular}}
\newcommand{\bcl}{\begin{center}}
\newcommand{\ecl}{\end{center}}
\newcommand{\bbt}{\bibitem}
\newcommand{\beq}{\begin{equation}}
\newcommand{\eeq}{\end{equation}}
\newcommand{\beqr}{\begin{eqnarray}}
\newcommand{\eeqr}{\end{eqnarray}}
\begin{document}
\normalsize
\parskip=5pt plus 1pt minus 1pt
\title{\boldmath Measurements of baryon pair decays of $\chicJ$ mesons }
\author{
M.~Ablikim$^{1}$, M.~N.~Achasov$^{6}$, O.~Albayrak$^{3}$, D.~J.~Ambrose$^{39}$, F.~F.~An$^{1}$, Q.~An$^{40}$, J.~Z.~Bai$^{1}$, Y.~Ban$^{26}$, J.~Becker$^{2}$, J.~V.~Bennett$^{16}$
% ...
\\
\vspace{0.2cm}
(BESIII Collaboration)\\
\vspace{0.2cm} {\it
$^{1}$ Institute of High Energy Physics, Beijing 100049, People's Republic of China\\
$^{2}$ Bochum Ruhr-University, D-44780 Bochum, Germany\\
$^{3}$ Carnegie Mellon University, Pittsburgh, Pennsylvania 15213, USA\\
% ...
$^{6}$ G.I. Budker Institute of Nuclear Physics SB RAS (BINP), Novosibirsk 630090, Russia\\
% ...
$^{16}$ Indiana University, Bloomington, Indiana 47405, USA\\
% ...
$^{26}$ Peking University, Beijing 100871, People's Republic of China\\
% ...
$^{39}$ University of Rochester, Rochester, New York 14627, USA\\
$^{40}$ University of Science and Technology of China, Hefei 230026, People's Republic of China\\
% ...
}}
\vspace{1.4cm}
\begin{abstract}
  Using 106 $\times 10^{6}$ $\psi^{\prime}$ decays collected with the BESIII
  detector at the BEPCII, three decays of $\chi_{cJ}$ ($J=0,1,2$) with
  baryon pairs ($\llb$, $\ssb$, $\SSB$) in the final state have been
  studied. The branching fractions are measured to be
  $\cal{B}$$(\chi_{c0,1,2}\rightarrow\Lambda\bar\Lambda) =(33.3 \pm 2.0 \pm
  2.6)\times 10^{-5}$, $(12.2 \pm 1.1 \pm 1.1)\times 10^{-5}$, $(20.8 \pm
  1.6 \pm 2.3)\times 10^{-5}$;
  $\cal{B}$$(\chi_{c0,1,2}\rightarrow\Sigma^{0}\bar\Sigma^{0})$ = $(47.8 \pm
  3.4 \pm 3.9)\times 10^{-5}$, $(3.8 \pm 1.0 \pm 0.5)\times 10^{-5}$,
  $(4.0 \pm 1.1 \pm 0.5) \times 10^{-5}$; and
  $\cal{B}$$(\chi_{c0,1,2}\rightarrow\Sigma^{+}\bar\Sigma^{-})$ = $(45.4 \pm
  4.2 \pm 3.0)\times 10^{-5}$, $(5.4 \pm 1.5 \pm 0.5)\times 10^{-5}$,
  $(4.9 \pm 1.9 \pm 0.7)\times 10^{-5}$, where the first error is
  statistical and the second is systematic. Upper limits on the
  branching fractions for the decays of
  $\chi_{c1,2}\rightarrow\Sigma^{0}\bar\Sigma^{0}$,
  $\Sigma^{+}\bar\Sigma^{-}$, are estimated to be
  $\cal{B}$$(\chi_{c1}\rightarrow\Sigma^{0}\bar\Sigma^{0}) < 6.2\times 10^{-5}$,
  $\cal{B}$$(\chi_{c2}\rightarrow\Sigma^{0}\bar\Sigma^{0}) < 6.5\times 10^{-5}$,
  $\cal{B}$$(\chi_{c1}\rightarrow\Sigma^{+}\bar\Sigma^{-}) < 8.7\times 10^{-5}$
  and $\cal{B}$$(\chi_{c2}\rightarrow\Sigma^{+}\bar\Sigma^{-}) < 8.8\times
  10^{-5}$ at the 90\% confidence level.
\end{abstract}
\maketitle
\linenumbers
\section{Introduction}
In the standard quark model, $\chicJ$ ($J$ = 0, 1, 2) mesons are
$\ccb$ states in an $L = 1$ configuration. Experimental studies on
$\chicJ$ decay properties are essential to test perturbative quantum
chromodynamics (QCD) models and QCD-based calculations.  The
importance of the color octet mechanism  for $\chicJ$ decays has
been pointed out for many years~\cite{Bodwin}, and theoretical
predictions of two-body exclusive decays have been made based on it.
The predictions of the color octet mechanism theory for some $\chicJ$ decays into baryon
pairs ($\BB$) disagree with measured values. For example, the branching fraction of
$\chi_{c0}\ar\llb$ is predicted to be $(93.5 \pm 20.5) \times 10^{-5}$ according to Ref.~\cite{production} and $(11.9 \sim 15.1)\times 10^{-5}$ according to Ref.~\cite{XHLiu}, while the world average of experimental measurements is $(33.0 \pm 4.0)\times 10^{-5}$~\cite{PDG}. One finds that the theoretical prediction is either about two times larger, or several times smaller than the experimental measurement. Although some
experimental results on $\chicJ$ exclusive decays have been
reported~\cite{bes02,bes03,cleo}, many decay modes of $\chicJ\ar\BB$
have not been observed yet, such as $\chi_{c1,2}\ar\ssb$, $\SSB$, or measured with poor
precision. For further testing of the color octet mechanism in the decays of the {\it P}-wave
charmonia, measurements of other baryon pair decays of $\chicJ$, such
as $\chicJ\ar\llb$, $\ssb$ and $\SSB$, are desired.

In addition, measurements of $\chi_{c0}\ar\BB$ are helpful  for further
understanding the helicity selection rule~\cite{hsr}, which
prohibits $\chi_{c0}$ decays into baryon-antibaryon pairs. However, the measured branching fractions for $\chi_{c0}\ar\BB$ do not vanish, for example $\chi_{c0}\ar\ppb$~\cite{PDG}, which demonstrates a strong
violation of the helicity selection rule in charmonium decays. It is necessary to
measure the decays of $\chi_{c0}\ar\BB$ in other channels to provide additional tests of
the helicity selection rule.

While $\chicJ$ mesons are not produced directly in $\EE$
annihilations, the large branching fractions of $\psp\ar\gamma\chicJ$
make $\EE$ collision at the $\psp$ peak a very clean environment
for $\chicJ$ investigation.  In this paper, the results of two-body
decays of $\chicJ\ar\llb$, $\ssb$ and $\SSB$ final states are
presented. This analysis is based on 106 $\times 10^{6}$ $\psp$ events~\cite{wangzy} collected with BESIII at the
BEPCII. A sample of 44 pb$^{-1}$ of data taken at $\sqrt{s}$ = 3.65 GeV
is used for continuum background study.
\section{BESIII DETECTOR AND MONTE CARLO SIMULATION}
BEPCII is a double-ring $\EE$ collider that has reached peak luminosity of about $0.6 \times 10^{33}~\rm{cm}^{-2}\rm{s}^{-1}$
at the peak energy of $\psi(3770)$.
The cylindrical
core of the BESIII detector consists of a helium-based main drift
chamber (MDC), a plastic scintillator time-of-flight system, and
a CsI(Tl) electromagnetic calorimeter (EMC), which are all enclosed in
a superconducting solenoidal magnet providing a 1.0 T magnetic
field. The solenoid is supported by an octagonal flux-return yoke with
resistive plate counter muon identifier modules interleaved with
steel. The acceptance for charged particles and photons is 93\% over
4$\pi$ stereo angle, and the charged-particle momentum and photon
energy resolutions at 1 GeV are 0.5\% and 2.5\%, respectively. The
detector is described in more detail in Ref.~\cite{BESIII}.

The BESIII detector is modeled with a Monte Carlo (MC) simulation
based on \textsc{geant}{\footnotesize 4}~\cite{geant4, geant42}.  The
$\psi^{\prime}$ resonance is produced with \textsc{kkmc} \cite{kkmc},
while the subsequent decays are generated with
\textsc{evtgen}~\cite{evt2} according to the branching fractions
provided by the Particle Data Group (PDG)~\cite{PDG}, and the
remaining unmeasured decay modes are generated with
\textsc{lundcharm}~\cite{lund}.
\section{Event selection}
The investigated final states include $\Lambda(\lamb)$, $p(\bar{p})$,
neutral $\pi^0$ mesons and a radiative photon from the decay
$\psp\rightarrow\gamma\chi_{cJ}$, where $\Lambda$($\lamb$) decays to
$\pi^-p$($\pi^+\pbar$), while $\pi^0$ is reconstructed in the decay to
$\pi^{0}\ar\GG$. Candidate events are required to satisfy the following
selection criteria.
A charged track should have good quality in the track fitting
and be within the angle coverage of the MDC ($|\cos\theta|<0.92$).
Photons are reconstructed from isolated showers in the EMC. The energy deposited
in the nearby TOF counter is included to improve the reconstruction
efficiency and energy resolution. Photon energies are required to be
greater than 25 MeV in the EMC barrel region ($|\cos\theta|<0.8$) and
greater than 50 MeV in the EMC end cap
($0.86 < |\cos\theta| < 0.92$). The showers in the angular range between
the barrel and the end cap are poorly reconstructed and excluded from
the analysis. Moreover, the EMC timing of the photon candidate must be
in coincidence with collision events, $0\leq t\leq 700$ ns, to
suppress electronic noise and energy deposits unrelated to the events.
\subsection{\boldmath $\chicJ\ar\llb$}
Candidate events contain at least two positively charged tracks, two
negatively charged tracks and one photon. The $\Lambda(\lamb)$
candidates are reconstructed from pairs of oppositely charged tracks,
which are constrained to secondary vertices and have invariant masses
closest to the nominal $\Lambda$ mass. The $\chi^{2}$ of the
secondary vertex fit must be less than 500. The candidate photon and the $\llb$ pair
are subjected to a four
constraint (4C) kinematic fit under the hypothesis of
$\psp\ar\gamma\llb$ to reduce background and improve the mass
resolution. When additional photons are found in an event, all
possible combinations are iterated over, and the one with the best
kinematic fit $\chi^2_{4C}$ is kept. Furthermore, $\chi^2_{4C}<50$ is
required to suppress potential background from $\psp\ar\ssb$. The
$\chi^2_{4C}$ selection criterion is determined by optimizing the
figure of merit (FOM), FOM = $\frac{S}{\sqrt{S + B}}$, where S is the
number of signal events and $B$ is the
number of background events based on the MC simulation.
Figure~\ref{scatterplot}(a) shows the comparison of $\chi^2_{4C}$
between data and MC simulation, which is normalized with the number of events satisfying the $\chi^{2}$ requirement. Figure~\ref{scatterplot}(b) shows
the scatter plots of $M_{p\pi^-}$ versus $M_{\bar{p}\pi^+}$ from the data.
Clear $\Lambda\bar\Lambda$ signals can be seen. The square around the $\Lambda$
nominal mass with a width of 20 MeV/$c^2$ is taken as the signal
region, which is also determined by maximizing the FOM.
From events with two or more photons, additional selection criteria are applied to
suppress backgrounds from $\ssb$ decays. The $\psp\ar\ssb$ candidates are
selected by minimizing
$\sqrt{(M_{\gamma\Lambda}-M_{\Sigma^0})^{2}+(M_{\gamma\lamb}-M_{\bar\Sigma^0})^{2}}$
from all combinations. However, some backgrounds remain in the signal
region from $\psp\ar\ssb$ events in which one photon from the
$\Sigma^{0}$ decays is not reconstructed. To remove these, events
falling into $|M_{\gamma\Lambda}-M_{\Sigma^0}| < 6$ MeV/$c^2$ and
$|M_{\gamma\lamb}-M_{\bar\Sigma^0}| <$ 6 MeV/$c^2$ have been discarded.
\subsection{\boldmath $\chicJ\ar\ssb$}
Candidate events have at least two positively charged tracks, two negatively charged tracks and three photons.
The charged track selection and $\Lambda(\bar\Lambda)$ reconstruction are the same as
described above for the $\chi_{cJ}\ar\llb$ decay. The mass window of
$\Lambda(\bar\Lambda)$ is optimized to be $|M_{p\pi}-M_{\Lambda}| <$ 7
MeV/$c^2$. The candidate photons and the $\llb$ pair
 are subjected to a 4C kinematic
fit under the hypothesis of $\psp\ar\gamma\GG\llb$ to reduce
background and improve the mass resolution. When additional photons
are found in an event, all possible combinations are looped over, the
one with the smallest $\chi^2_{4C}$ is kept, and
$\chi^2_{4C} < 35$ is required to suppress the dominant background from
$\psp\ar\ssb$.  Figure~\ref{scatterplot}(c) shows the comparison of
$\chi^2_{4C}$ between data and MC simulation, which is normalized with the number of events satisfying the $\chi^{2}$ requirement. The $\ssb$ candidates
are chosen by minimizing
$\sqrt{(M_{\gamma\Lambda}-M_{\Sigma^0})^{2}+(M_{\gamma\lamb}-M_{\bar\Sigma^0})^{2}}$. Figure~\ref{scatterplot}(d)
shows the scatter plot of $M_{\gamma\Lambda}$ versus
$M_{\gamma\lamb}$ from the data. Clear $\ssb$ signals can be seen. The square
around the $\Sigma^{0}$ nominal mass with a width of 32 MeV/$c^2$ represents the signal region.
\subsection{\boldmath $\chicJ\ar\SSB$}
\begin{figure}[hbt]
\bcl
\subfigure{\includegraphics[width=0.4\textwidth]{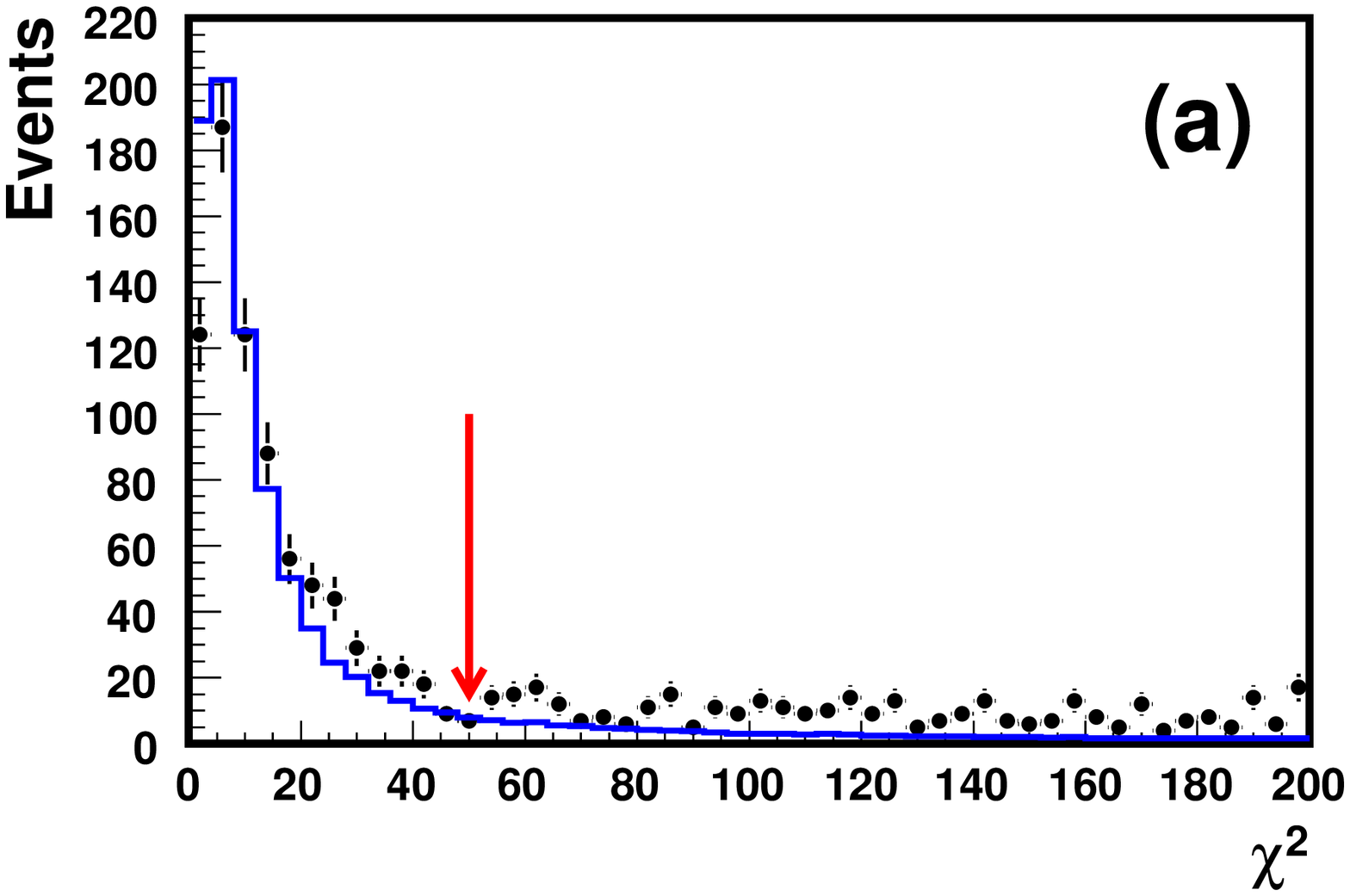}}
\subfigure{\includegraphics[width=0.4\textwidth]{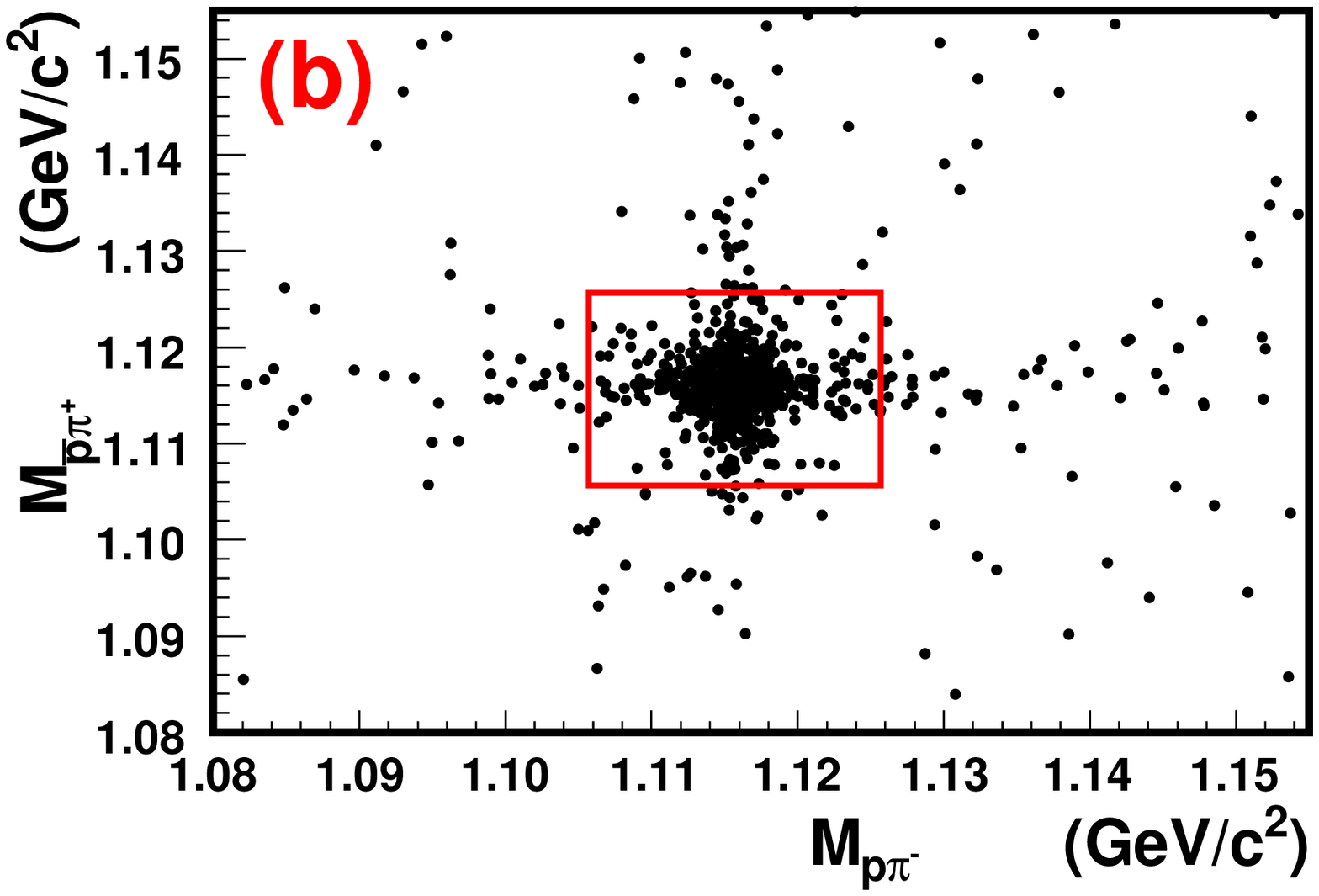}}
\subfigure{\includegraphics[width=0.4\textwidth]{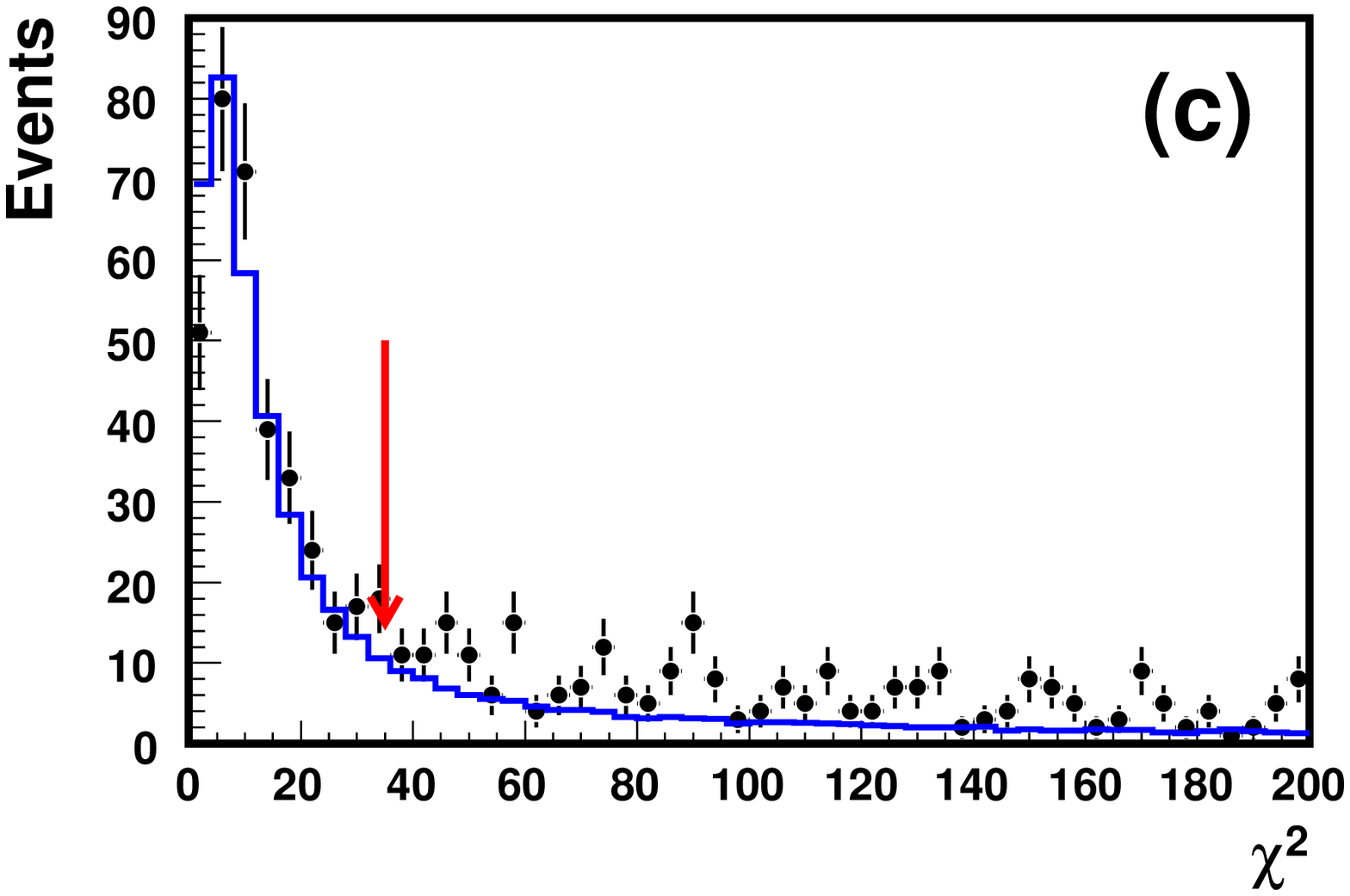}}
\subfigure{\includegraphics[width=0.4\textwidth]{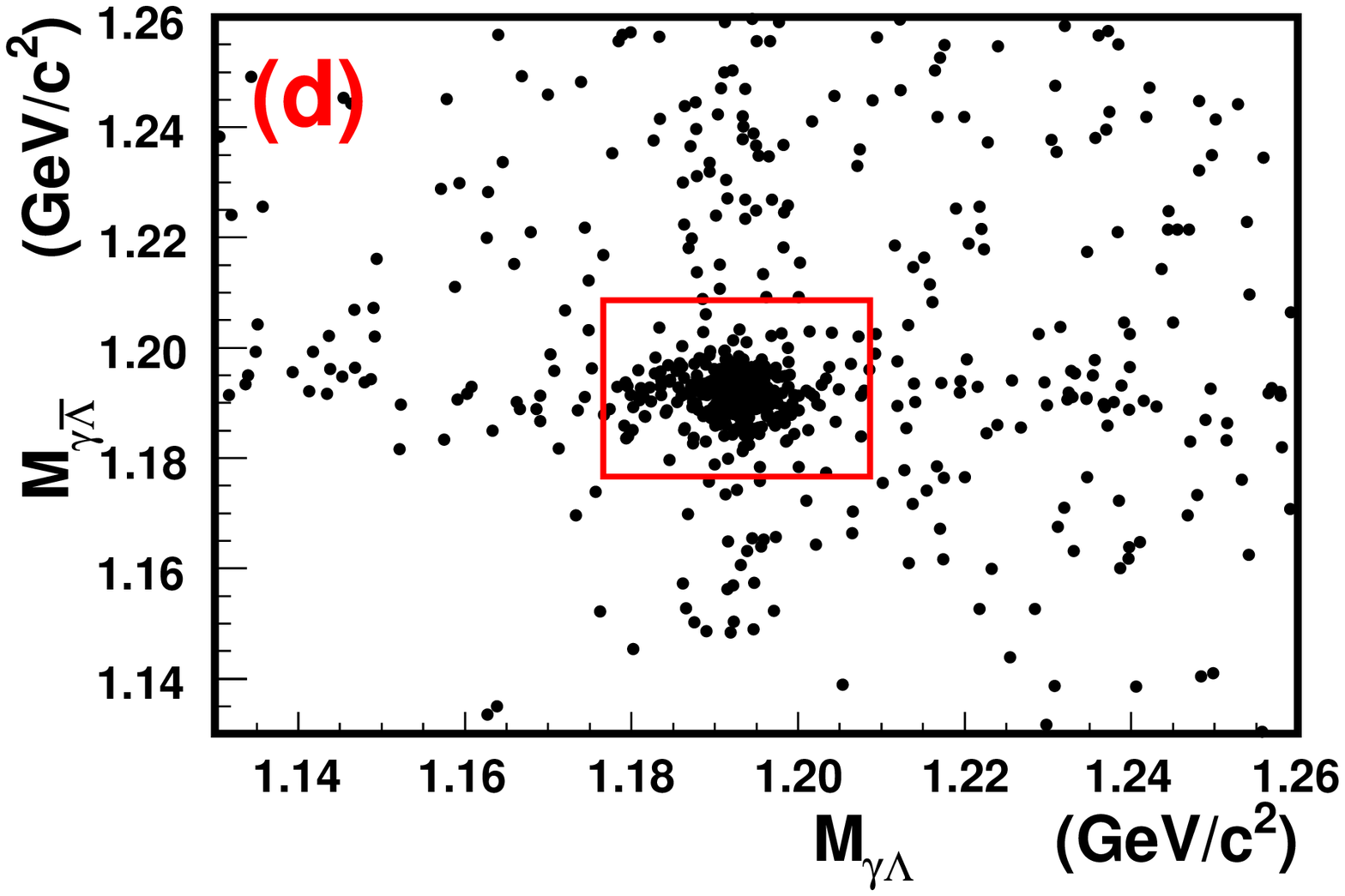}}
\subfigure{\includegraphics[width=0.4\textwidth]{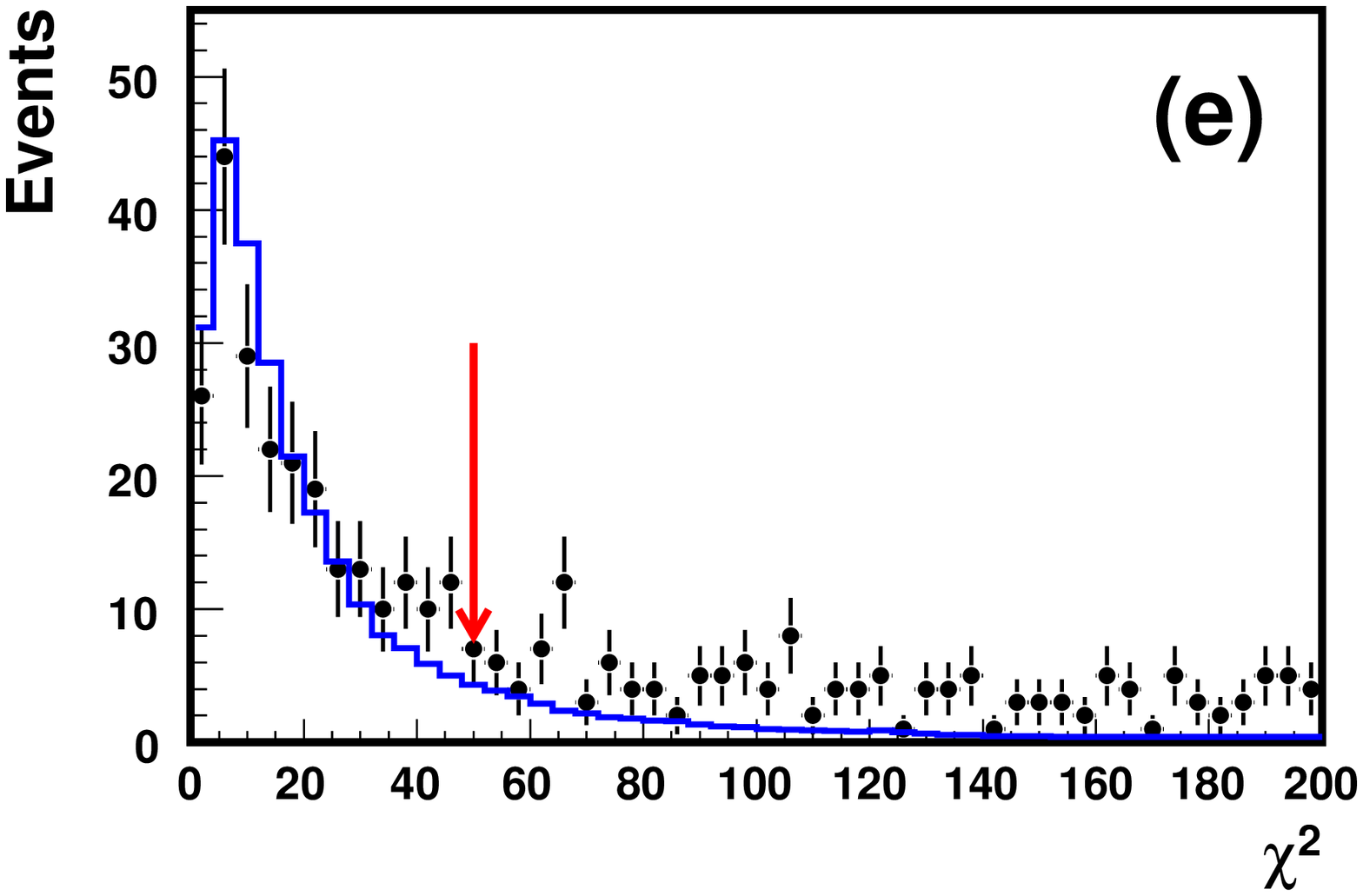}}
\subfigure{\includegraphics[width=0.4\textwidth]{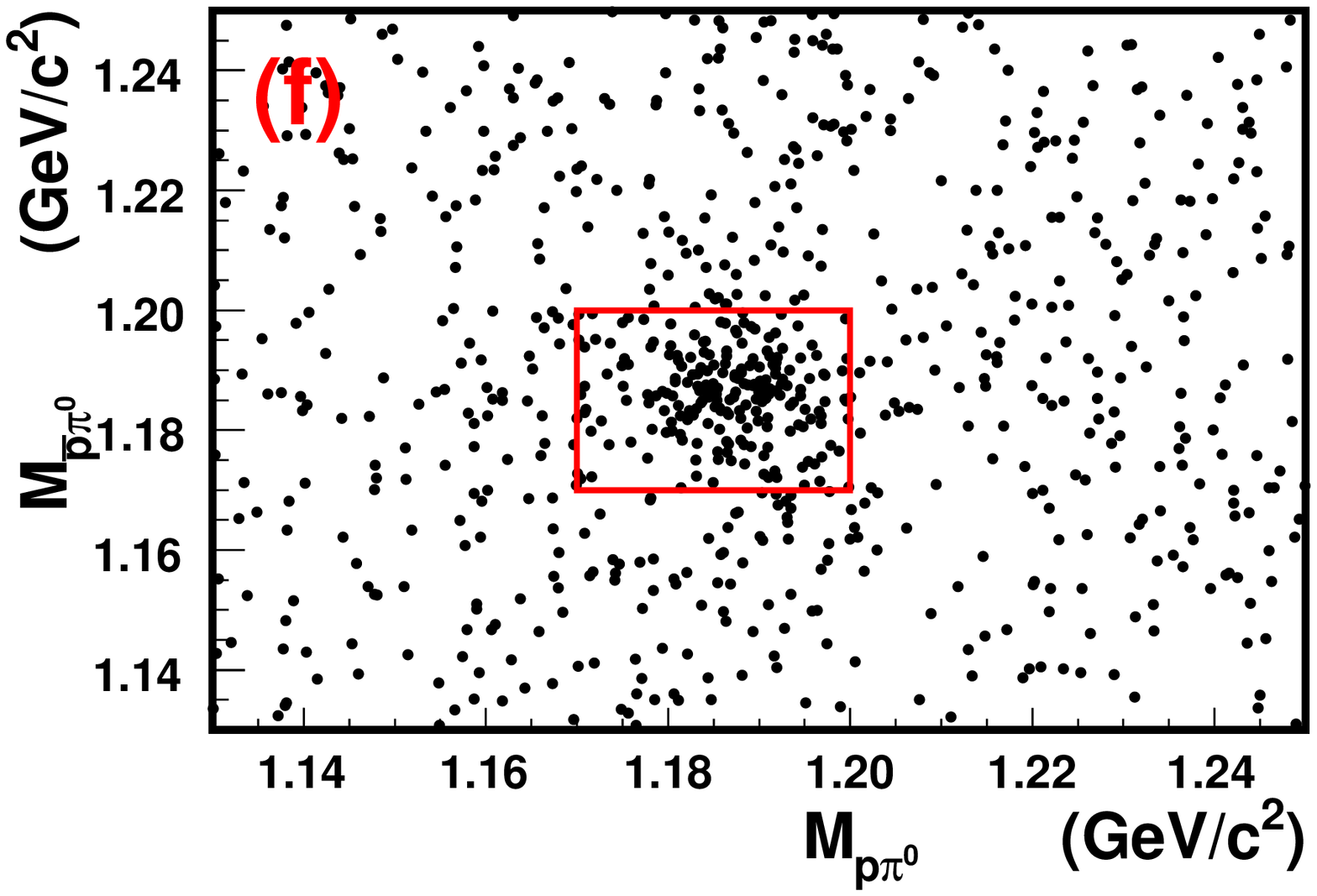}}
\caption{
(a) The $\chi^2_{4C}$ distribution and (b) $M_{p\pi^-}$ versus $M_{\bar{p}\pi^+}$ (data) for the $\psp\ar\gamma\chicJ, \chicJ\ar\llb$ candidates; (c) the $\chi^2_{4C}$ distribution and (d) $M_{\gamma\Lambda}$ versus $M_{\gamma\lamb}$ (data) for the $\psp\ar\gamma\chicJ, \chicJ\ar\ssb$ candidates;
(e) the $\chi^2_{4C}$ distribution and (f) $M_{p\pi^0}$ versus $M_{\bar{p}\pi^0}$ (data) for the $\psp\ar\gamma\chicJ, \chicJ\ar\SSB$
candidates.}
\label{scatterplot}
\ecl
\end{figure}

Candidate events contain at least one positively charged, one
negatively charged tracks and five photons. We impose a 4C kinematic fit to the selected tracks and photons under the
$\psp\ar 5\gamma\ppb$ hypothesis and keep the one with the smallest $\chi^2_{4C}$, and $\chi^2_{4C} < 50$ is required to suppress
the dominant background from $\psp\ar\SSB$. Figure~\ref{scatterplot}(e)
shows the comparison of $\chi^2_{4C}$ between data and MC simulation, which is normalized with the number of events satisfying the $\chi^{2}$ requirement.
The $\pi^0$ candidates are reconstructed by selecting the combination which
minimizes
$\sqrt{(M_{\GG}^{(1)}-M_{\pi^0})^{2}+(M_{\GG}^{(2)}-M_{\pi^0})^{2}}$. The
$\SSB$ pair is selected by minimizing $\sqrt{(M_{p\pi^0}-
  M_{\Sigma^+})^{2}+(M_{\bar{p}\pi^0}-
  M_{\bar{\Sigma}^-})^{2}}$. Figure~\ref{scatterplot}(f) shows the
scatter plot of $M_{p\pi^0}$ versus $M_{\bar{p}\pi^0}$ from the data. Clear $\SSB$ signals can be seen. The square of 1.17 GeV/$c^2$
$< M_{p\pi^0} <$ 1.20 GeV/$c^2$ and 1.17 GeV/$c^2$$ < M_{\bar{p}\pi^0} <$
1.20 GeV/$c^2$ denotes the signal region.
\section{Background study}
\subsection{Continuum backgrounds}
The events collected at E$_{\rm cm} = 3.65$ GeV, whose integrated luminosity
is more than 1/4 of $\psp$ samples, are analyzed to estimate the contribution from the continuum process. No events are survived in the $\llb$,
$\ssb$ and $\SSB$ signal regions. Therefore, backgrounds from the
continuum are neglected.
\subsection{\boldmath Dominant backgrounds in $\llb$, $\ssb$ and $\SSB$ final states }
By using 106 $\times 10^{6}$ inclusive MC events, we find that
the dominant background for $\chicJ\ar\llb$ comes from the decay
$\psp\ar\ssb$ in which one photon is missing. The non-$\llb$
background from the decay $\chicJ\ar\pppr$ is
negligibly small due to the low efficiency near the mass threshold.
For $\chicJ\ar\ssb$, the dominant background is also found to
arise from $\psp\ar\ssb$. But this background mainly distributes around the $\psi^{\prime}$ mass region in the $\ssb$ invariant mass.
In addition, a few background events come from $\psp \ar\ppjpsi$ and
$\psp\ar\xxb$. For $\chicJ\ar\SSB$, the backgrounds are small; they are from the decay $\psp\ar\SSB$, $\psp\ar\ppjpsi$ and
$\jpsi\ar\ppb$ (or $\gamma\ppb$).
The contributions of all backgrounds mentioned above are estimated by MC simulation
according to their branching fractions.
\section{\boldmath Fit to the signal of $\chicJ$}
\begin{figure}
\bcl
\subfigure{\includegraphics[width=0.6\textwidth]{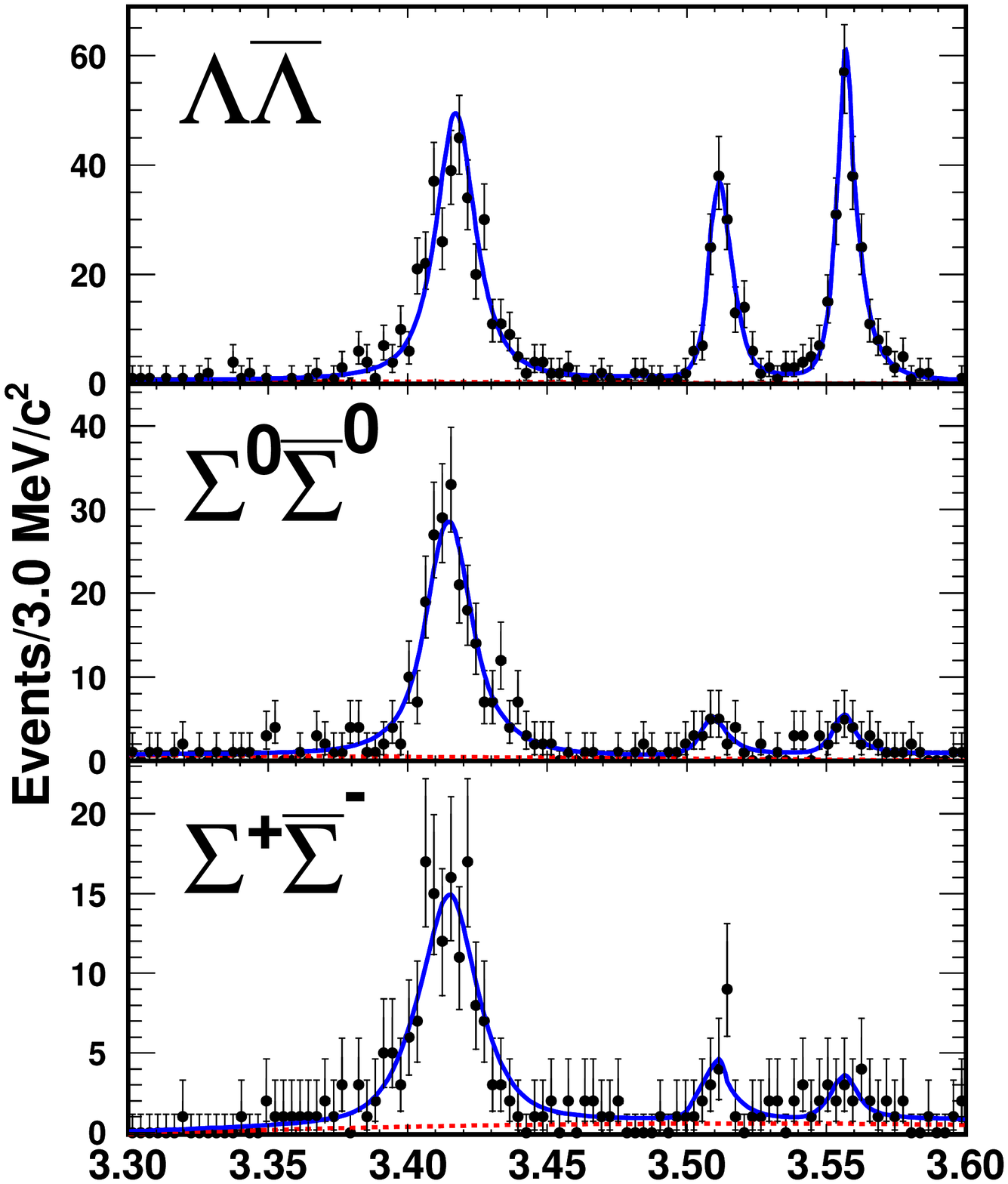}}
\put(-192,0){\Large \textbf{M$_{B\bar{B}}$}~~~~~(GeV/$c^{2}$)}
\caption{The fit to the invariant mass M$_{\BB}$. The dots with error bars are
  for data. The solid line is the fit results. Dashed-line is
  other background. The parameters of signal function are fixed to those obtained from MC simulation.}
\label{fit}
\ecl
\end{figure}

The invariant mass of the baryon pairs M$_{B\bar{B}}$ for all selected events
are shown in Figs.~\ref{fit}(a)--(b) for
$\chicJ\ar\llb,~\ssb$ and $\SSB$, respectively.  Clear $\chi_{c0,1,2}$
signals can be seen in $\llb$ final state, and a clear $\chi_{c0}$ signal is seen
in both $\ssb$ and $\SSB$ final states, while the $\chi_{c1,2}$ signals are not significant in $\ssb$ and $\SSB$ final states. We fit the invariant mass spectra of
baryon pairs, M$_{\BB}$, to extract the numbers of $\chicJ$ signal
events, where the signals are represented by Breit-Wigner functions
convolved with a Crystal Ball function to account for the detector
resolution, a second-order Chebychev polynomial is used to describe
non-peaking backgrounds, and the dominant background events, estimated
by MC simulation, have been directly subtracted from the data. The widths
of the Breit-Wigner functions were fixed according to the
known values~\cite{PDG}, the parameters of the Crystal Ball function are fixed based on MC simulation, and these parameters are
varied by $\pm$ $\sigma$ for the determination of systematic
uncertainties. To determine the goodness of fit, we bin the data so that the number of events in each bin is at least ten. The calculated $\chi^2$/d.o.f is 1.03, 1.53 and 1.71 for the $\llb$, $\ssb$ and $\SSB$ final states, respectively. The numbers of
$\chi_{c0,1,2}$ signal events from the fits are listed in
Table~\ref{number00}.
For the decay $\chi_{c1,2}\ar\ssb$, $\SSB$, the upper limits of the branching fractions at the 90\% C.L. are also determined
with a Bayesian method~\cite{bayes}. The statistical significances of
the signals are calculated as $\sqrt{-2\Delta \ln{\cal L}}$, where $\Delta
\ln{\cal L}$ is the difference between the logarithmic maximum
likelihood values of the fit with and without the corresponding
signal function. They are 4.3$\sigma$ and 4.6$\sigma$ for
$\chi_{c1,2}\ar\ssb$, and 4.4$\sigma$ and 3.0$\sigma$ for
$\chi_{c1,2}\ar\SSB$, respectively.
The signal efficiencies determined from MC simulation are also listed in Table~\ref{number00}, where
the proper angular distributions for photons emitted in
$\psp\ar\gamma\chicJ$ are used~\cite{E1}. The decay of $\chicJ\ar\BB$
and the decay of baryons are generated with a phase space model.
\begin{table}[!h]
  \centering
  {\caption{Efficiencies ($\epsilon$ in~$\%$) obtained from MC simulation,
      and the signal yields N$^{obs}$ determined from fit.}\label{number00}}
\begin{tabular}{ccccccc}  \hline \hline
\multicolumn{1}{c}{} &\multicolumn{2}{c}{$\chi_{c0}$} &\multicolumn{2}{c}{$\chi_{c1}$} &\multicolumn{2}{c}{$\chi_{c2}$} \\ \hline
 Mode                       &N$^{obs}$     &$\epsilon$    &N$^{obs}$    &$\epsilon$        &N$^{obs}$     &$\epsilon$  \\ \hline
$\Lambda\bar\Lambda$        &368.9 $\pm$ 22.1 &26.6 $\pm$ 0.2 &135.6 $\pm$ 12.6 &27.9 $\pm$ 0.2  &207.1$\pm$15.7  &26.3 $\pm$ 0.2   \\
$\Sigma^{0}\bar{\Sigma^{0}}$&242.8 $\pm$ 17.1 &12.2 $\pm$ 0.1 &20.0 $\pm$ 5.3   &13.2 $\pm$ 0.1 &18.9 $\pm$ 5.3   &12.7 $\pm$ 0.1   \\
$\Sigma^{+}\bar\Sigma^{-}$  &147.8 $\pm$ 13.8 &12.3 $\pm$ 0.1 &18.0 $\pm$ 5.4    &13.1 $\pm$ 0.1 &14.5 $\pm$ 5.6    &12.3 $\pm$ 0.1   \\ \hline \hline
\end{tabular}
\end{table}
\section{Systematic error}
The systematic errors mainly originate from the uncertainties of the tracking
efficiency, $\Lambda(\lamb)$ reconstruction efficiency, the photon efficiency, 4C kinematic fit, the branching fractions of the intermediate states, fit range, the angular distribution of $\chi_{c1,2}\ar\BB$, background shape,
signal line shape, MC resolution and the total number of $\psp$ events.
\begin{enumerate}
\item The decay $\psp\ar\llb$ with $\Lambda\ar p\pi^{-}$ and $\Lambda\ar\bar{p}\pi^{-}$ is employed to study the
 $\Lambda(\lamb)$ reconstruction efficiency.
The selection criteria of charged tracks are the same as before except we use
  particle identification information to suppress background. Candidate events have at
  least one positively charged and one negatively charged tracks, which are required to be identified as a $\pi^+(\pi^{-})$ track and an $\bar{p}(p)$ track, respectively. Also, the invariant mass of $\pi^+\bar{p}(\pi^{-}p)$ must be within 10 MeV/$c^2$ of the nominal $\lamb$ mass. Furthermore, the
  momentum of $\lamb(\Lambda)$ candidates is required to be within 20 MeV/$c$ of
  its nominal value in two-body decay of $\psp\ar\llb$.  The number of
  $\Lambda$ signal events, $N_{\Lambda}^0$, is extracted by fitting
  the recoiling mass spectrum of $\lamb$, $M_{recoil}^{\lamb}$. Then
  two additional oppositive charged tracks, a $\pi^-(\pi^{+})$ and a $p(\bar{p})$, are required
  to reconstruct $\Lambda$ and are constrained to the secondary vertex. The number of $\Lambda$ signal events,
  $N_{\Lambda}^1$, is extracted by fitting $M_{recoil}^{\lamb}$ after
  requiring a $\Lambda$ secondary vertex constraint. The $\Lambda(\bar\Lambda)$ reconstruction efficiency
  is determined as $\epsilon_{\Lambda} =\frac{N_{\Lambda}^1}{N_{\Lambda}^0}$. The difference of the efficiencies between data and MC simulation is found to be 2.0\% for a $\Lambda$ and 5.0\% for a $\lamb$, which are taken as the systematic error due to $\Lambda(\lamb)$ reconstruction efficiency.
\item  Since the decay length for $\Sigma^+(\bar\Sigma^-)$ is small, the decay $J/\psi\rightarrow\pi^{+}\pi^{-}p\bar{p}$ is used to study the MDC tracking efficiency for the proton and antiproton of the $\Sigma^{+}\bar\Sigma^{-}$ final state. It is found that the efficiency for MC simulated events agrees with that determined from data within 1.0\% for each charged track. Hence, 2.0\% is taken as the systematic error for the proton and antiproton of the $\Sigma^{+}\bar\Sigma^{-}$ final state.
\item The uncertainty due to photon detection efficiency is 1\% per
  photon, which is determined from the decay $J/\psi\rightarrow\rho\pi$~\cite{bianjm}.
\item Five decays, $\jpsi\ar\llb,\jpsi\ar\ssb,
  \jpsi\ar\Xi^0\bar{\Xi}^0, \psp\ar\ppjpsi$ ($\jpsi\ar\ppb$) and
  $\psp\ar\ppjpsi (\jpsi\ar\ppb\pi^0)$, are used to study the
  efficiencies of the 4C kinematic fits. The
  signal events are selected from data and inclusive MC events without the
  4C fit information. The remaining background is found to be
  negligible according to the studies of the inclusive MC
  events.
  The efficiency of the 4C kinematic fit is defined as $\frac{N_1}{N_0}$,
  where $N_{0}$ is the the number of signal events, and $N_{1}$ is the
  number of events survived. For the $\chi_{cJ}\ar\llb$, where the final state is $\psp\ar\gamma\llb$, two
  decays, $\jpsi\ar\llb$, and $\jpsi\ar\ssb$, are used to
  investigate the systematic error due to the 4C kinematic fit. The final states of these two
  control samples contain one photon less or more than the signal
  channel. Conservatively, the larger difference observed in the two
  control samples, 2.4\%, is taken as the systematic error. Similarly, the
  larger difference in $\jpsi\ar\ssb$ and $\jpsi\ar\xxb$, 2.9\%, is
  taken as the systematic error of the $\chicJ\ar\ssb$ channel, and the larger difference in
  $\psp\ar\ppjpsi ~(\jpsi\ar\ppb)$ and $\psp\ar\ppjpsi
  ~(\jpsi\ar\ppb\pi^0)$, 1.3\%, is taken as the error of
  $\chicJ\ar\SSB$.
\item When changing mass ranges in fitting M$_{\BB}$ signals
 to 3.30--3.62 GeV/$c^2$ or to 3.25--3.62 GeV/$c^2$, the fitted numbers of $\chi_{c0,1,2}$ have some changes for data and MC simulation. Taking the $\llb$ channel as an example, the results in the range of 3.30 GeV/$c^2$ to 3.60 GeV/$c^2$ are taken as central values, when the fit range is changed to 3.32--3.60 GeV/$c^2$, the changes relative to central values are found to be 2.7\%, 3.6\% and 2.2\% for the $\chi_{c0,1,2}$ decays, respectively, while in the range 3.25--3.62 GeV/$c^2$, the changes are found to be  2.2\%, 0.9\% and 4.3\%.
    Conservatively, we take the larger ones, 2.7\%, 3.6\% and 4.3\%, as the systematic errors for the $\llb$ final state.  With the same method, the systematic errors for the other two channels are determined to be 1.4\%, 6.7\% and 4.3\% for the $\ssb$ final state and 1.4\%, 3.0\% and 7.2\% for the $\SSB$ final state.
\item In the fits to the M$_{\BB}$ invariant mass, the signals are described by a parameterized shape obtained from MC simulation
    in which the  widths of $\chicJ$ are fixed since we only observe a small number of
  signal events in $\chi_{c1,2}\ar\ssb$ and $\SSB$. When changing the parameters of
  $\chicJ$ widths in this MC simulation by $\pm$ $\sigma$, it is found that the difference of
  the numbers of fitted $\chi_{c1,2}$ events between data and MC is
  1.2\%, 0.0\% and 0.0\% for the $\Lambda\bar\Lambda$ final state; 1.9\%, 0.0\% and 3.7\%
  for the $\Sigma^{0}\bar\Sigma^{0}$ final state and 1.0\%, 0.5\% and 2.0\% for the
  $\Sigma^{+}\bar\Sigma^{-}$ final state. Hence, we take the difference as the
  systematic error due to the $\chicJ$ widths.
\item The partial width for an E1/M1 radiative transition is
  proportional to the cube of the radiative photon energy
  ($E^{3}_{\gamma}$), which leads to a diverging tail in the lower mass
  region. Two damping factors have been proposed by the
  KEDR~\cite{KEDR} and the CLEO~\cite{CLEO01} Collaborations and have
  been included to describe the signal line shape. Differences in the signal yields with respect
  to the fit not taking into account this damping factor are
  observed, and the greater differences are 0.7\%, 2.1\% and 2.7\% for the $\llb$ final state; 1.4\%, 1.0\% and 2.2\% for the $\ssb$ final state; 0.0\%, 2.7\% and 5.5\% for the $\SSB$ final state, which are taken as the
  systematic error associated with the signal line shape.

\item From the decay $J/\psi\rightarrow\Lambda\bar\Lambda$,
it is found that the average resolution is 7.90 $\pm$ 0.09 MeV/$c^{2}$ for the data and 7.08 $\pm$ 0.04 MeV/$c^{2}$ for MC. Differences in fitting the $\chi_{cJ}$ signal with and without fixing the MC parameters are found to be 1.5\%, 0.5\% and 2.4\% for the $\Lambda\bar\Lambda$ final states, which are taken as the systematic error of the resolution. However, from the decays $J/\psi\rightarrow\Sigma^{0}\bar\Sigma^{0}$ and
  $J/\psi\rightarrow\Sigma^{+}\bar\Sigma^{-}$, one can find that the resolutions between data and MC are consistent.
  Therefore, the systematic errors of the resolution for the $\Sigma^{0}\bar\Sigma^{0}$ and $\Sigma^{+}\bar\Sigma^{-}$ final
  state are neglected.
\item To estimate the uncertainty of the angular distribution, we use
  another model in which the angular distribution of
  $\chi_{c1,2}\ar\BB$ is taken into account according to the helicity
  amplitude~\cite{Liao}. When the two independent helicity amplitudes, $B_{\frac{1}{2},-\frac{1}{2}}$ and $B_{-\frac{1}{2},\frac{1}{2}}$,  are set to be 1.0, the efficiencies are found to be (28.8 $\pm$ 0.2)\% and (27.9 $\pm$ 0.2)\% for the $\chi_{c1,2}\ar\llb$ final state, respectively. The differences from phase space are 3.2\% and 6.0\%. Similar comparisons are also done for the $\ssb$ and $\SSB$ final states, and the differences are smaller. Conservatively, we take the difference of the $\llb$ final state as the systematic error of the angular distribution for all $B\bar{B}$ final states.
\item In Fig.~\ref{fit}, the combinatorial background curves are
  fitted with a second-order Chebychev polynomial. The background
  function is changed to first- and third-order polynomials, and
  the largest difference is taken as the systematic error due to the
  uncertainty in the description of the background shape.
\item The total number of $\psp$ events are obtained by studying
  inclusive hadronic $\psp$ decays with an uncertainty of
  0.81\%~\cite{wangzy}.
\end{enumerate}
Table~\ref{error} lists all systematic error contributions, and the
total systematic error is obtained by adding the individual
contributions in quadrature.
\btbl[h]
\caption{Systematic errors in the branching fraction measurements (\%)}.
\bcl
\doublerulesep 2pt
\begin{tabular}{cccccccccc}  \hline \hline
\multicolumn{1}{l}{}  &\multicolumn{3}{c}{$\chi_{cJ}\rightarrow\Lambda\bar\Lambda$} &\multicolumn{3}{c}{$\chi_{cJ}\rightarrow\Sigma^0\bar\Sigma^0$} &\multicolumn{3}{c}{$\chi_{cJ}\rightarrow\Sigma^+\bar\Sigma^-$} \\ \hline
Source                           &$\chi_{c0}$ &$\chi_{c1}$ &$\chi_{c2}$ &$\chi_{c0}$ &$\chi_{c1}$ &$\chi_{c2}$ &$\chi_{c0}$ &$\chi_{c1}$ &$\chi_{c2}$  \\
The total number of $\psp$       &0.81 &0.81&0.81    &0.81 &0.81&0.81   &0.81 &0.81 &0.81   \\
MDC tracking~($p, \bar{p}$)       &--   &--  &--      &--   &--  &--     &2.0  &2.0  &2.0    \\
Photon efficiency                &1.0  &1.0 &1.0     &3.0  &3.0  &3.0   &5.0  &5.0  &5.0    \\
$\Lambda$ reconstruction         &2.0  &2.0 &2.0     &2.0  &2.0 &2.0    &--   &--   &--      \\
$\bar\Lambda$ reconstruction     &5.0  &5.0 &5.0     &5.0  &5.0 &5.0    &--   &--   &--      \\
Kinematic fit                    &2.4  &2.4 &2.4     &2.9  &2.9 &2.9    &1.3  &1.3  &1.3    \\
Fitting range                    &2.7  &3.6 &4.3     &1.4  &6.7 &4.3    &1.4  &3.0  &7.2    \\
$\chi_{cJ}$ width                &1.2  &0.0 &0.0     &1.9  &0.0  &3.7   &1.0  &0.5  &2.0    \\
Angular distribution             &0.0  &3.2 &6.0     &0.0   &3.2 &6.0   &0.0  &3.2  &6.0     \\
Background shape                 &0.5  &1.3 &1.3     &1.7  &7.8 &6.0    &1.8  &2.5  &3.0    \\
Signal line shape                 &0.7  &2.1 &2.7     &1.4  &1.0 &2.2    &0.0  &2.7  &5.5    \\
MC resolution                    &1.5  &0.5 &2.4     &0.0  &0.0 &0.0    &0.0  &0.0  &0.0     \\
$\cal{B}$($\psp\rightarrow\gamma\chi_{cJ}$)&3.2&4.3   &4.0   &3.2  &4.3 &4.0    &3.2  &4.3  &4.0     \\
$\cal{B}$($\Sigma\rightarrow p\pi$)   &--   &--  &--      &--   &--  &--     &0.82 &0.82  &0.82 \\
$\cal{B}$($\Lambda\rightarrow p\pi)$     &1.1 &1.1&1.1       &1.1&1.1&1.1    &--   &--  &--          \\
Total systematic error       &7.7 &9.3 &11.1     &8.3  &13.6 &13.2  &7.0 &9.1 &13.4   \\ \hline \hline
\end{tabular}
\label{error}
\ecl
\etbl
\section{Results}
The branching fraction of $\chicJ\ar\BB$ is determined by
\[
{\cal B}(\chicJ\ar\BB)=\frac{N^{obs}[\chicJ]}{N_{\psp}\cdot\epsilon \cdot\prod_{i}{\cal B}_{i}},
\]
and if the signal is not significant, the corresponding upper limit of branching fraction is set with
\[
{\cal B}(\chicJ\ar\BB)<\frac{N^{obs}_{UL}[\chicJ]}{N_{\psp}\cdot\epsilon\cdot\prod_{i}{\cal B}_{i}\cdot(1.0-\sigma_{sys})},
\]
where, $N^{obs}$ is the number of observed signal events and
$N_{UL}^{obs}$ is the upper limit of the number of events, $\epsilon$
is the detection efficiency shown in Table~\ref{number00},
$\sigma_{sys}$ is the relative the systematic error, $N_{\psp}$ is the
total number of $\psp$ events~\cite{wangzy}, and $\prod_{i}{\cal
  B}_{i}$ is the product of the branching fractions taken from the world average~\cite{PDG}
  for the $\psp\rightarrow\gamma\chicJ$ and the other decays
that are involved.  With the numbers listed in Table~\ref{number00}
and the branching fractions for the relevant baryon decays, the
branching fractions or the upper limits at the 90\% C.L. for $\chicJ$
decays are determined, as listed in Table~\ref{result}.  \btbl[h]
\caption{Branching fractions (or their upper limits) of $\chicJ\ar\llb,\ssb$ and $\SSB$ (in units of $ 10^{-5}$). The first error is statistical and the second is systematic.}
\bcl
\footnotesize{
\doublerulesep 2pt
\begin{tabular}{c|c|c|c|c}  \hline \hline
Mode                  &          &$\chi_{c0}$                  &$\chi_{c1}$                 &$\chi_{c2}$\\ \hline
                      &This work &33.3 $\pm$ 2.0 $\pm$ 2.6         &12.2 $\pm$ 1.1 $\pm$ 1.1        &20.8 $\pm$ 1.6 $\pm$ 2.3  \\
                      &PDG       &33.0 $\pm$ 4.0                 &11.8 $\pm$ 1.9                &18.6 $\pm$ 2.7     \\
$\Lambda\bar\Lambda$  &CLEO     &33.8 $\pm$ 3.6 $\pm$ 2.2 $\pm$ 1.7&11.6 $\pm$ 1.8 $\pm$ 0.7 $\pm$ 0.7&17.0 $\pm$ 2.2 $\pm$ 1.1 $\pm$ 1.1    \\
                      &Theory &($93.5 \pm 20.5^{a}$, $22.1 \pm 6.1^{b}$) $^{\textsc{\cite{production}}}$  &--          &($15.2 \pm 1.7^{a}$, $4.3 \pm 0.6^{b}$) $^{\textsc{\cite{production}}}$  \\
                      &      &11.9 $\sim$ 15.1 $^{ \textsc{\cite{XHLiu}}}$  &3.9 $^{ \textsc{\cite{wong}}}$&3.5 $^{ \textsc{\cite{wong}}}$   \\ \hline
                      &This work &47.8 $\pm$ 3.4 $\pm$ 3.9         &3.8 $\pm$ 1.0 $\pm$ 0.5 \color{red}{($ < $~6.2)}
                      &4.0 $\pm$ 1.1 $\pm$ 0.5 \color{red}{($< $~6.5)} \\
                      &PDG       &42.0 $\pm$ 7.0                 & $<$ 4.0                      &$<$ 8.0  \\
$\Sigma^0\bar\Sigma^0$&CLEO      &44.1 $\pm$ 5.6 $\pm$ 4.2 $\pm$ 2.2 &$<$ 4.4                      &$<$ 7.5   \\
                      &Theory &($25.1 \pm 3.4^{a}$, $18.7 \pm 4.5^{b}$) $^{\textsc{\cite{production}}}$  &--         &($38.9 \pm 8.8^{a}$, $4.2 \pm 0.5^{b}$) $^{\textsc{\cite{production}}}$  \\
                      &   &--                     &3.3 $^{ \textsc{\cite{wong}}}$                         &5.0 $^{ \textsc{\cite{wong}}}$   \\ \hline
                      &This work &45.4 $\pm$ 4.2 $\pm$ 3.0         &5.4 $\pm$ 1.5 $\pm$ 0.5 \color{red}{($<$~8.7)}
                      &4.9 $\pm$ 1.9 $\pm$ 0.7 \color{red}{($<$ 8.8)}\\
                      &PDG       &31.0 $\pm$ 7.0                 &$<$ 6.0                       &$<$ 7.0   \\
$\Sigma^+\bar\Sigma^-$&CLEO     &32.5 $\pm$ 5.7 $\pm$ 4.0 $\pm$ 1.7 &$<$ 6.5                       &$<$ 6.7  \\
                      &Theory    &5.5 $\sim$ 6.9 $^{ \textsc{\cite{XHLiu}}}$  &3.3 $^{ \textsc{\cite{wong}}}$  &5.0 $^{ \textsc{\cite{wong}}}$   \\
\hline \hline
\end{tabular}}
\label{result}
\ecl
\etbl
\section{Summary}
Three $\chicJ$ decays to the baryon pairs are observed, and their branching fractions are measured at BESIII,
which are consistent with the world averages within the errors. For the decay
of $\chicJ\ar\llb$, the experimental results are still inconsistent with
theoretical predictions~\cite{wong, XHLiu,production}, which are helpful to check the theoretical model of decays of $\chicJ\ar\llb$.
For the decays of $\chi_{c1,2}\ar\ssb$ and $\SSB$, the significances are
improved relative to the previous measurments, but the comparisons of their branching fractions
between experiments and theoretical predictions are inconclusive due
to the limited experimental precision.
\section{acknowledgement}
The BESIII collaboration thanks the staff of BEPCII and the computing center for their hard efforts. This work is supported in part by the Ministry of Science and Technology of China under Contract No. 2009CB825200, 2009CB825206; National Natural Science Foundation of China (NSFC) under Contracts Nos. 10625524, 10821063, 10825524, 10835001, 10935007, 10975143, 10975047, 10979008, 11125525, 11275057; Joint Funds of the National Natural Science Foundation of China under Contracts Nos. 11079008, 11079027, 11179007; the Chinese Academy of Sciences (CAS) Large-Scale Scientific Facility Program; CAS under Contracts Nos. KJCX2-YW-N29, KJCX2-YW-N45; 100 Talents Program of CAS; Istituto Nazionale di Fisica Nucleare, Italy; Ministry of Development of Turkey under Contract No. DPT2006K-120470; U. S. Department of Energy under Contracts Nos. DE-FG02-04ER41291, DE-FG02-91ER40682, DE-FG02-94ER40823; U.S. National Science Foundation; University of Groningen (RuG) and the Helmholtzzentrum fuer Schwerionenforschung GmbH (GSI), Darmstadt; WCU Program of National Research Foundation of Korea under Contract No. R32-2008-000-10155-0.

\end{document}